  \documentclass[12pt]{article}
  \usepackage{times}

  \topmargin - 0.5 cm 
\begin{document}

% \baselineskip = 0.50 true cm
  \baselineskip = 13 pt
\begin{center}
{\large \bf ANGULAR MOMENTUM AND MUTUALLY UNBIASED BASES}  
\end{center}

\vspace{0.5cm}

\begin{center}
{\bf MAURICE R.~KIBLER}
\end{center}

\begin{center}
{Institut de physique nucl\'eaire de Lyon}\\
{IN2P3-CNRS/Universit\'e Claude Bernard Lyon 1}\\
{43 bd du 11 novembre 1918}\\
{F-69622 Villeurbanne Cedex, France}\\
{kibler@ipnl.in2p3.fr}
\end{center}

 \vspace{2.5cm}

  \noindent {\bf Abstract}

The Lie algebra of the group SU$_2$ is constructed from two
deformed oscillator algebras for which the deformation 
parameter is a root of unity. This leads to an unusual 
quantization scheme, the $\{ J^2 , U_r \}$ scheme, an 
alternative to the familiar $\{ J^2 , J_z \}$ quantization 
scheme corresponding to common eigenvectors of the Casimir 
operator $J^2$ and the Cartan operator $J_z$. A connection 
is established between the eigenvectors of the complete set 
of commuting operators $\{ J^2 , U_r \}$ and mutually 
unbiased bases in spaces of constant angular momentum.

 \vspace{2.5cm}

Key words: angular momentum; deformations; harmonic oscillator;
Lie algebra; polar decomposition; MUBs.

 \vspace{2.5cm}

To be published in International Journal of Modern Physics B

   \newpage

\section{Introduction}

In recent years, the notion of deformed oscillator algebra and its 
extension to deformed Lie algebra, or Hopf algebra in mathematical 
parlance,$^{1-5}$ proved to be useful in various fields of theoretical 
physics. For instance, one- and two-parameter deformations of oscillator 
algebras and Lie algebras were successfully applied to statistical 
mechanics$^{6-13}$ and to nuclear, atomic and molecular physics.$^{14-18}$
In the case where the deformation parameter is a root of unity, let us 
also mention the importance of deformed oscillator algebras for the 
definition of $k$-fermions, which are objects interpolating between 
fermions and bosons,$^{19}$ and the study of fractional 
supersymmetry.$^{20}$  

The aim of this note is two-fold. First, we show 
how a deformation of two truncated harmonic oscillators 
leads to a polar decomposition of the Lie algebra of SU$_2$. 
Such a decomposition is especially appropriate  
for developing the representation theory and the 
Wigner--Racah algebra of SU$_2$ in a non-standard 
basis adapted to cyclical symmetry.$^{21}$ Second, we establish a 
contact between the corresponding bases for spaces of constant 
angular momentum and the so-called mutually unbiased bases (MUBs) 
in a finite-dimensional Hilbert space. The latter bases$^{22-44}$ 
play a central role in quantum information theory. In particular, 
the use of quantum-mechanical states belonging to MUBs is of 
paramount importance in quantum cryptography (securing quantum 
key exchange) and quantum state tomography (deciphering a quantum state). 

\section{Angular Momentum Theory in a Nonstandard Basis}

\subsection{The Lie algebra of SU$_2$ from two oscillator algebras}
Let ${\cal F}(1)$ and 
    ${\cal F}(2)$ be two finite-dimensional 
Hilbert spaces of dimension $k$ with $k \in {\bf N} \setminus \{ 0 , 1\}$. 
We use $( \ | \ )$ to denote the inner product on ${\cal F}(i)$ and, 
for each space ${\cal F}(i)$ with $i=1,2$, 
we choose an orthonormal basis 
$\{ | n_i ) : n_i = 0, 1, \cdots, k-1 \}$. Let 
$( a_{i-}, a_{i+}, N_i )$ be a triplet of linear operators on 
${\cal F}(i)$ defined by 
  \begin{eqnarray}
  a_{i\pm} |n_i) = 
  \left( \left[ n_i + s \pm \frac{1}{2} \right]_q \right)^{\alpha_{i\pm}} |n_i \pm 1) 
  \nonumber \end{eqnarray}
  \begin{eqnarray}
  a_{i+} |k-1) = 0, \quad a_{i-} |0) = 0, \quad 
%  \nonumber \end{eqnarray}
%  \begin{eqnarray}
  N_i |n_i) = n_i |n_i)
  \nonumber \end{eqnarray}
where 
\begin{eqnarray}
s = \frac{1}{2}, \quad \alpha_{i\pm} = \frac{1 \pm (-1)^i}{2}, \quad
% \nonumber \end{eqnarray}
% and
% \begin{eqnarray}
     q = \exp \left( {2 \pi {\rm i} \over k} \right), \quad  
     \left[ x \right]_q = \frac{1-q^x}{1-q},  \quad  x \in {\bf R} 
\nonumber \end{eqnarray}
with $i=1,2$. It can be shown that the operators $a_{i-}$, $a_{i+}$ and $N_i$ 
satisfy the following relations 
   \begin{eqnarray}
   a_{i-}a_{i+} - qa_{i+}a_{i-} = 1,            \quad  
   \left( a_{i\pm} \right)^k = 0,               \quad 
   \left[ N_i, a_{i\pm} \right] = \pm a_{i\pm}, \quad
   N_i^{\dagger} = N_i
%  \nonumber 
   \end{eqnarray}
where we use the notation $A^{\dagger}$ for the adjoint of $A$ and 
$\left[ A, B \right]$ for the commutator of the operators $A$ and $B$. 
The two algebras defined by Eq.~(1) with $i=1,2$ are two commuting oscillator algebras 
with $q$ being a root of unity; this
is reminiscent of the two oscillator algebras used for the introduction 
of $k$-fermions.$^{19, 20}$

We now consider the space ${\cal F}_k = {\cal F}(1) \otimes {\cal F}(2)$ 
of dimension $k^2$. An orthonormal basis for ${\cal F}_k$ is provided by the 
vectors 
\begin{eqnarray}
  | n_1 , n_2 ) = | n_1 ) \otimes | n_2 ), 
  \quad n_i = 0, 1, \cdots, k-1, 
  \quad i = 1,2
\nonumber \end{eqnarray}
The key of our derivation of a nonstandard basis of SU$_2$ 
consists in defining the two linear operators
  \begin{eqnarray}
  H = {\sqrt {N_1 \left( N_2 + 1 \right) }},  \quad U_r = s_{1+} s_{2-}
  \nonumber \end{eqnarray}
where 
  \begin{eqnarray}
  s_{i \pm}  =  a_{i\pm} + {\rm e}^{\frac{1}{2} {\rm i} \phi_r}  {1 \over 
  \left[ k-1 \right]_q!} (a_{i\mp})^{k-1} 
  \nonumber \end{eqnarray}
for $i=1,2$. In the operator $s_{i \pm}$, the phase $\phi_r$ 
is an arbitrary real parameter taken in the form 
  \begin{eqnarray}
  \phi_r = \pi (k-1) r,  \quad  r \in {\bf R}
  \nonumber \end{eqnarray}
and $\left[ n \right]_q!$ stands for the $q$-deformed factorial 
defined by 
  \begin{eqnarray}
  \forall n \in {\bf N}^* : 
  \left[ n \right]_q! = 
  \left[ 1 \right]_q 
  \left[ 2 \right]_q \cdots 
  \left[ n \right]_q, \quad 
  \left[ 0 \right]_q! = 1  
  \nonumber \end{eqnarray}  
It is immediate to show that the action of $H$ and $U_r$ on 
${\cal F}_k$ is given by
  \begin{eqnarray}
  H |n_1 , n_2) = {\sqrt{ n_1 (n_2 + 1) } |n_1 , n_2)},
  \quad n_i = 0, 1, 2, \cdots, k-1, 
  \quad i=1,2
  \nonumber \end{eqnarray}
and
  \begin{eqnarray}
  U_r |n_1 , n_2) = |n_1+1 , n_2-1), 
                  \quad n_1 \not = k-1,  
                  \quad n_2 \not = 0
  \nonumber \end{eqnarray}
  \begin{eqnarray}
  U_r |k-1 , n_2) = {\rm e}^{\frac{1}{2} {\rm i} {\phi}_r} 
  |0 , n_2 - 1),   \quad n_2 \not= 0
  \nonumber \end{eqnarray}
  \begin{eqnarray}
  U_r |n_1 , 0) = {\rm e}^{\frac{1}{2} {\rm i} {\phi}_r} 
  |n_1 + 1 , k-1), \quad n_1 \not= k-1  
  \nonumber \end{eqnarray}
  \begin{eqnarray}
  U_r |k-1 , 0) = {\rm e}^{{\rm i} {\phi}_r} 
  |0 , k-1)
  \nonumber \end{eqnarray}
The operators $H$ and $U_r$ satisfy interesting properties. 
The operator $H$ is Hermitean and the operator 
$U_r$ is unitary. Furthermore, the action of $U_r$ on the space ${\cal F}_k$
is cyclic in the sense that
\begin{eqnarray}
(U_r)^k = {\rm e}^{{\rm i} {\phi}_r} I
\nonumber \end{eqnarray}
where $I$ is the identity operator.

From the Schwinger work on angular momentum,$^{45}$
we introduce 
  \begin{eqnarray}
  J = {1 \over 2} \left( n_1+n_2 \right),  \quad  
  M = {1 \over 2} \left( n_1-n_2 \right)  
  \nonumber \end{eqnarray}
We shall use the notation 
  \begin{eqnarray} 
  |J M \rangle \equiv |J + M , J-M) = |n_1 , n_2)
  \nonumber \end{eqnarray}
For a fixed value of $J$,
the label $M$ can take $2J+1$ values $M = -J, -J+1, \cdots, J$. For 
fixed $k$, the maximum value of $J$ is 
$J = J_{\rm max} = k-1$ and the following value of $J$
\begin{eqnarray}
J = j = \frac{1}{2} (k-1)
\nonumber \end{eqnarray}
is admissible. For a given value of $k \in {\bf N} \setminus \{0 , 1\}$, 
the $2j+1=k$ vectors $|j m \rangle$ belong to the vector space 
${\cal F}_k$. Let $\varepsilon (j)$ be the subspace of ${\cal F}_k$, 
of dimension $\dim \varepsilon (j) = k$, spanned by the $k$ vectors 
$|j m \rangle$ with $m = -j, -j+1, \cdots, j$. We can thus associate 
the space $\varepsilon (j)$ for $j = \frac{1}{2}, 1, \frac{3}{2}, \cdots$ 
to the values $k = 2, 3, 4, \cdots$, respectively.  The subspace 
$\varepsilon (j)$ of ${\cal F}_k$ is stable under $H$ and $U_r$. Indeed,
    the action of the operators $H$ and $U_r$ on the space $\varepsilon (j)$
can be described by 
  \begin{eqnarray}
  H |j m \rangle = {\sqrt{ (j+m)(j-m+1) }} |j m \rangle
  \nonumber \end{eqnarray} 
and
  \begin{eqnarray}
  U_r |jm \rangle = \left[ 1 - \delta (m,j) \right] |j m+1 \rangle + \delta(m,j)  
  {\rm e}^{{\rm i} {\phi}_r}
  |j -j \rangle 
  \nonumber \end{eqnarray}
We can check that the operator $H$ is Hermitean and the operator $U_r$ is 
unitary on the space $\varepsilon (j)$. Furthermore, we have 
$\left( U_r \right)^{2j+1} = {\rm e}^{ {\rm i} {\phi}_r } I$
which reflects the cyclical character of $U_r$ on $\varepsilon (j)$. 

We are now in a position to give a realization of the Lie algebra of the group 
SU$_2$ in terms of $U_r$, $N_1$ and $N_2$. Let us define the 
three operators
  \begin{eqnarray}
  J_+ = H           U_r,  \quad  
  J_- = U_r^{\dagger} H,  \quad
  J_z = {1 \over 2} \left( N_1-N_2 \right)
  \nonumber 
  \end{eqnarray}
It is straightforward  to  check that 
  \begin{eqnarray}
  J_{\pm} |j m \rangle = {\sqrt{ (j \mp m)(j \pm m+1) }} |j m \pm 1 \rangle,
  \quad
  J_z   |j m \rangle = m |jm \rangle
  \nonumber \end{eqnarray}
Consequently, we get the commutation relations 
  \begin{eqnarray}
  \left[ J_z , J_{\pm} \right] = \pm J_{\pm},  \quad 
  \left[ J_+ , J_-     \right] = 2J_z 
  \nonumber \end{eqnarray}
which correspond to the Lie algebra of SU$_2$. 

\subsection{An nonstandard basis for the group SU$_2$}
The decomposition of the shift operators $J_+$ and $J_-$ in terms 
of $H$ and $U_r$ coincides with the polar
decomposition worked out in Refs.~46 and 47 
in a completely different 
way. This is easily seen by taking the matrix elements of
$U_r$ and $H$ in the $\{ J^2 , J_z \}$ quantization scheme  
and by comparing these elements to the ones of the operators 
$\Upsilon$ and $J_T$ in Ref.~46. We 
are thus left with 
$H = J_T$ and, by identifying the arbitrary phase $\varphi$ 
of Ref.~46 with $\phi_r$, we 
obtain $U_r = \Upsilon$
so that $J_+ = J_T \Upsilon$ and $J_- = \Upsilon ^{\dagger} J_T$. 

It is immediate to check that the Casimir operator 
$J^2$ of  the Lie algebra su$_2$ can be rewritten as
\begin{eqnarray}
J^2 = \frac{1}{4} (N_1 + N_2) (N_1 + N_2 + 2) 
\nonumber \end{eqnarray}
in terms of $N_1$ and $N_2$. It is a simple matter of calculation
to prove that $J^2$ commutes with $U_r$ for any value of $r$. Therefore, 
for $r$ fixed,
the commuting set $\{ J^2, U_r\}$ provides us with an alternative to the
familiar commuting set $\{ J^2, J_z \}$ of angular momentum theory.

The eigenvalues and the common eigenvectors 
of the complete set of commuting operators 
$\{ J^2, U_r \}$ can be easily found. This 
leads to the following result.

\vspace{\baselineskip}

\noindent {\bf Result 1}. The spectra of the operators 
$U_r$ and $J^2$ are given by
  \begin{eqnarray}
  U_r | j n_{\alpha} ; r \rangle = q^{-\alpha} 
      | j n_{\alpha} ; r \rangle, \quad
  J^2 | j n_{\alpha} ; r \rangle = j(j+1) 
      | j n_{\alpha} ; r \rangle 
  \nonumber \end{eqnarray}
where 
  \begin{eqnarray}
  |j n_{\alpha} ; r \rangle = {1 \over {\sqrt{2j + 1}}}
  \sum_{m = -j}^j
  q^{\alpha m} 
  |j m \rangle, \quad q = \exp \left( {\rm i} \frac{2 \pi}{2 j + 1}   \right) 
% \nonumber 
  \end{eqnarray}
with the range of values 
  \begin{eqnarray}
  \alpha = - jr + n_{\alpha}, \quad n_{\alpha} = 0, 1, \cdots, 2j 
  \nonumber \end{eqnarray}
where $2j \in {\bf N}^*$ and $r \in {\bf R}$.

\vspace{\baselineskip}

Each vector $| j n_{\alpha} ; r \rangle$ can be 
considered as a discrete Fourier transform$^{47}$ 
in the finite-dimensional Hilbert space $\varepsilon(j)$.
As a  matter  of fact, the inter-basis expansion coefficients 
\begin{eqnarray}
\langle jm | j n_{\alpha} ; r \rangle = 
{1 \over \sqrt{2 j + 1}} 
\exp \left[ {\rm i} \frac{2 \pi}{2 j + 1} (- jr + n_{\alpha}) m \right]
\nonumber \end{eqnarray}
(with $m   = -j, -j  + 1, \cdots,  j$ 
and 
$n_{\alpha}=  0,       1, \cdots, 2j$) in Eq.~(2) 
define a unitary transformation, in 
$\varepsilon(j)$ (with $j = \frac{1}{2}, 1, \frac{3}{2}, \cdots$), 
that allows to pass from the well-known 
orthonormal standard spherical basis 
\begin{eqnarray}
s(j)   = \{ |j m \rangle :  m = - j, - j + 1, \cdots, j \}
\nonumber \end{eqnarray}
to the orthonormal non-standard basis
\begin{eqnarray}
b_r(j) = \{ |j n_{\alpha} ; r \rangle : n_{\alpha} = 0, 1, \cdots, 2j \}
\nonumber \end{eqnarray}

For a given value of $r$, the basis $b_r(j)$ 
is an alternative to the spherical basis
$s(j)$ of the space $\varepsilon(j)$. Two bases 
$b_r(j)$ and $b_s(j)$ with $r \not=s$ are thus 
two equally admissible orthonormal bases for 
$\varepsilon(j)$. The state vectors of the bases 
$b_r(j)$ and $b_s(j)$ are common eigenstates of 
$\{ J^2 , U_r \}$ and $\{ J^2 , U_s \}$, 
respectively. The overlap between the
bases $b_r(j)$ and $b_s(j)$ is controlled by
     \begin{eqnarray}
     \langle j n_{\alpha} ; r | j n_{\beta} ; s \rangle = \frac{1}{2j+1} 
     \frac{\sin (\alpha - \beta) \pi}
     {\sin (\alpha - \beta) \frac{\pi}{2j+1}}
     \nonumber \end{eqnarray}
with $\alpha = - jr + n_{\alpha}$ and 
     $\beta  = - js + n_{\beta }$ 
where $n_{\alpha}, n_{\beta } = 0, 1, \cdots, 2j$.

\section{Connection with Mutually Unbiased Bases}

We are now ready for establishing contact with MUBs. Let 
$\varepsilon(d)$ be a Hilbert space of dimension $d$ endowed 
with the inner product $\langle \ | \ \rangle$. Two 
orthonormal bases 
$A = \{ |A \alpha \rangle : \alpha = 0, 1, \cdots, d-1 \}$ and 
$B = \{ |B \beta  \rangle : \beta  = 0, 1, \cdots, d-1 \}$ are 
said to be mutually unbiased if and only if 
$\vert \langle A \alpha | B \beta \rangle \vert = \frac{1}{\sqrt{d}}$
for all $\alpha \in \{ 0, 1, \cdots, d-1 \}$ and 
    all $\beta  \in \{ 0, 1, \cdots, d-1 \}$. For an  
arbitrary value of $d$, the number of MUBs cannot be 
greater than $d+1$.$^{22-25}$ 

We note  in  passing that  the latter  result  can  be  justified 
from group theory. The  $d$  orthonormal vectors of a basis for 
$\varepsilon(d)$  can  be  considered  as  a basis for a fundamental 
representation of dimension $d$ of the group SU$_d$. This group is 
of dimension  $d^2 - 1$  and of rank  (i.e.,  the  number of Cartan 
generators) $d - 1$. Therefore, the maximal number of independent 
sets of $d-1$ 
commuting operators it is possible to construct from the  $d^2 - 1$ 
generators  of  SU$_d$  is  $\frac{d^2 -1}{d-1} = d+1$.   This  is 
precisely the maximum number of MUBs for the space $\varepsilon(d)$. 
Indeed, the limit $d+1$ is reached if $d$ is a prime number$^{23}$  
or a power of a prime number.$^{24-30}$

It is also interesting to note that a connection 
exists between MUBs and various geometries 
(e.g., see Refs.~31, 36, 38, 40 and 43). In particular, 
according to the SPR conjecture,$^{31}$ for $d$ fixed with 
$d$ not equal to a power of a prime number, the problem 
of the existence of a complete set of $d+1$ MUBs 
would be equivalent to the one of the existence 
of projective planes of order $d$. 

We derive below some preliminary results of interest for 
an investigation of a relation between the $\{ J^2, U_r\}$ 
scheme and MUBs. To begin with, from Eq.~(2), we have 
the following result.

\vspace{\baselineskip}

\noindent {\bf Result 2.} The overlap between the
bases $s(j)$ and $b_r(j)$ satifies
\begin{eqnarray}
| \langle jm | j n_{\alpha} ; r \rangle |^2 = {1 \over \dim {\varepsilon(j)}} 
\nonumber \end{eqnarray}
so that $s(j)$ and $b_r(j)$ are two MUBs for the space 
$\varepsilon(j)$.  

\vspace{\baselineskip}

As an illustration, we consider the space 
$\varepsilon(\frac{1}{2})$ of dimension 2. 
Equation~(2) yields
   \begin{eqnarray}
   |\frac{1}{2} 0 ; 0 \rangle = \frac{1      }{\sqrt{2}} \left( 
   |\frac{1}{2} -\frac{1}{2} \rangle + 
   |\frac{1}{2}  \frac{1}{2} \rangle \right), \quad 
   |\frac{1}{2} 1 ; 0 \rangle = \frac{{\rm i}}{\sqrt{2}} \left( 
 - |\frac{1}{2} -\frac{1}{2} \rangle 
 + |\frac{1}{2}  \frac{1}{2} \rangle \right)
   \nonumber \end{eqnarray}
for $r = 0$ and
   \begin{eqnarray}
   |\frac{1}{2} 0 ; 1 \rangle = \frac{1}{\sqrt{2}} \left( 
   \rho      |\frac{1}{2} -\frac{1}{2} \rangle + 
   \rho^{-1} |\frac{1}{2}  \frac{1}{2} \rangle \right), \quad 
   |\frac{1}{2} 1 ; 1 \rangle = \frac{1}{\sqrt{2}} \left( 
   \rho^{-1} |\frac{1}{2} -\frac{1}{2} \rangle + 
   \rho      |\frac{1}{2}  \frac{1}{2} \rangle \right)
   \nonumber \end{eqnarray}
for $r=1$ with $\rho = {\rm e}^{{\rm i} \frac{\pi}{4}}$. It 
is evident that the three bases 
$s(\frac{1}{2})$, $b_0(\frac{1}{2})$ and $b_1(\frac{1}{2})$ 
constitute a complete 
set of MUBs for $\varepsilon(\frac{1}{2})$. 

The situation is not so simple for 
$2j \in {\bf N} \setminus \{ 0 , 1\}$. For fixed $j$, 
the eigenfunctions of the operators $U_r$ and $U_s$, with 
$r \not= s$, are not necessarily independent. We give in 
what follows some results that can be useful for $2 j \not= 1$.

\vspace{\baselineskip}

\noindent {\bf Result 3.} By assuming 
     \begin{eqnarray}
     s = r + \frac{n_{\beta} - n_{\alpha}}{j} + 
     \frac{2j + 1}{j} k_{\alpha \beta}, \quad
     k_{\alpha \beta} \in {\bf Z}
%    \nonumber 
     \end{eqnarray}
we get
     \begin{eqnarray}
     | j n_{\beta}  ; s \rangle = (-1)^{2 j k_{\alpha \beta}} 
     | j n_{\alpha} ; r \rangle  
     \nonumber \end{eqnarray}
and the corresponding bases $b_r(j)$ and $b_s(j)$ are not MUBs. 

\vspace{\baselineskip}

\noindent {\bf Result 4.} The commutator of $U_s$ and $U_r$ on $\varepsilon(j)$ 
assumes the form 
     \begin{eqnarray}
     \left[ U_s , U_r \right] = \left( {\rm e}^{ {\rm i} {\phi}_s } - 
                          {\rm e}^{ {\rm i} {\phi}_r } \right)
     \left[	| j , - j   \rangle \langle j , j - 1 | - 
     	| j , - j+1 \rangle \langle j , j     | \right]
     \nonumber \end{eqnarray}
Therefore, a necessary and sufficient condition that the operators 
$U_s$ and $U_r$ commute is 
     \begin{eqnarray}
     s = r + \frac{x}{j}, \quad
     x \in {\bf Z}
%    \nonumber 
     \end{eqnarray}
We note that Eq.~(3) implies Eq.~(4).

\vspace{\baselineskip}

\noindent {\bf Result 5.} On the space $\varepsilon(j)$, let $Z$ be the familiar 
phase operator defined by 
     \begin{eqnarray}
     \forall m \in \{ -j, -j+1, \cdots, j \} : 
     Z | j m \rangle = q^{-m} | j m \rangle
     \nonumber \end{eqnarray}
and, for fixed $r$, let $V_{ra}$ be the $2j+1$ unitary operators given by
     \begin{eqnarray}
     V_{ra} = U_r Z^a = q^{a} Z^a U_r, \quad a = 0, 1, \cdots, 2j 
     \nonumber \end{eqnarray}
(cf. the Weyl commutation relation rule). The Hilbert-Schmidt inner product of the operators $V_{sb}$ and $V_{ra}$
is 
     \begin{eqnarray}
     {\rm tr} \left( V_{sb}^{\dagger} V_{ra} \right) = (2j+1) \delta(a, b) + 
     q^{j(b-a)} \left[ {\rm e}^{ {\rm i} ({\phi}_r  - {\phi}_s )} -1 \right]
     \nonumber \end{eqnarray}
where the trace is taken on $\varepsilon(j)$ and 
where $r \in {\bf R}$, $s \in {\bf R}$ and $a,b = 0, 1, \cdots, 2j$. 
For $r$ and $s$ such that 
the condition (4) is satisfied, we have 
      \begin{eqnarray}
     {\rm tr} \left( V_{sb}^{\dagger} V_{ra} \right) = (2j+1) \delta(a, b) 
     \nonumber \end{eqnarray}
with $r \in {\bf R}$, $s \in {\bf R}$ and $a,b = 0, 1, \cdots, 2j$. 

\vspace{\baselineskip}
   
We note that $2j = 1$ is the sole case for which it is possible 
to find $r$ and $s$ such that ${\rm tr} \left( U_{s}^{\dagger} U_{r} \right) = 0$. This 
explains the peculiarity of the case $2j = 1$.   

\vspace{\baselineskip}

\noindent {\bf Result 6.} In the case where $2j+1$ is prime, 
following the works in Refs.~26, 27, 35 and 47, 
for a given value of $r$ let $M$ be the set of 
unitary operators 
     \begin{eqnarray}
     M = \{ V_{ra} : a = 0, 1, \cdots, 2j \}
     \nonumber \end{eqnarray}
generated by the two generalized Weyl-Pauli operators $U_r$ and $Z$. The vectors
of the spherical basis $s(j)$ and the eigenvectors 
of the $2j+1$ operators in $M$ provide a set of
of $2j+2$ MUBs for the Hilbert space $\varepsilon(j)$ of dimension $2j+1$. 

\vspace{\baselineskip}

The derivation of the latter result easily follows by adapting the 
proof of Theorem 2.3 of Ref.~26.

As an example, we treat the case $j=1$ with $r=0$ for which the 12 vectors of
the 4 MUBs can be described by a single simple formula. The $2j+2 = 4$ MUBs 
consist of the spherical basis $s(1)$ and of the 3 bases (corresponding to 
$a = 0$, 1 and 2) spanned by the vectors 
     \begin{eqnarray}
     \Psi_a(n_{\alpha}) = \frac{1}{\sqrt{3}} 
     \left( \omega^{- n_{\alpha} + a}| 1 -1 \rangle 
           +                         | 1  0 \rangle 
           +\omega^{  n_{\alpha} + a}| 1  1 \rangle \right)
     \nonumber \end{eqnarray}
with $n_{\alpha} = 0$, 1, 2 and $a = 0$, 1, 2 (as usual, 
$\omega = {\rm e}^{{\rm i} \frac{2 \pi}{3}}$).  The vectors 
$| 1 m \rangle$ of the spherical basis $s(1)$ are eigenvectors 
of $J_z$ with the real eigenvalues $m$. The case $a=0$ 
corresponds to the basis $b_0(1)$, the vectors $| 1 n_{\alpha} ; 0 \rangle$
of which are eigenvectors 
of $V_{00} = U_0$ with the complex eigenvalues $\omega^{- n_{\alpha}}$. More 
generally, for fixed $a$ (with $a = 0$, 1 or 2), 
the vectors of the basis $\{ \Psi_a(n_{\alpha}) : n_{\alpha} = 0, 1 \ {\rm and} \ 2 \}$ 
are eigenvectors of the operators $V_{0a}$:
     \begin{eqnarray}
     V_{0a}\Psi_a(n_{\alpha}) = \omega^{- n_{\alpha} - a} \Psi_a(n_{\alpha})
     \nonumber \end{eqnarray}
As a r\'esum\'e, by introducing the notation 
     \begin{eqnarray}
     N_{xyz} (x, y, z) \equiv 
     N_{xyz} 
     \left( x | 1 -1 \rangle + y | 1 0 \rangle + z | 1 1 \rangle \right), 
     \quad N_{xyz} = \frac{1}{\sqrt{|x|^2 + |y|^2 + |z|^2}}
     \nonumber \end{eqnarray}
we have the 4 MUBs
     \begin{eqnarray}
     s(1)  &:& (1, 0, 0) ; (0, 1, 0) ; (0, 0, 1)   \nonumber \\
     a = 0 &:& \frac{1}{\sqrt{3}} (1, 1, 1) ; 
               \frac{1}{\sqrt{3}} (\omega^2, 1, \omega  ) ;
	       \frac{1}{\sqrt{3}} (\omega  , 1, \omega^2) \nonumber \\
     a = 1 &:& \frac{1}{\sqrt{3}} (\omega  , 1, \omega  ) ; 
               \frac{1}{\sqrt{3}} (       1, 1, \omega^2) ;
	       \frac{1}{\sqrt{3}} (\omega^2, 1, 1       ) \nonumber \\
     a = 2 &:& \frac{1}{\sqrt{3}} (\omega^2, 1, \omega^2) ; 
               \frac{1}{\sqrt{3}} (\omega  , 1, 1       ) ;
	       \frac{1}{\sqrt{3}} (       1, 1, \omega  ) \nonumber 
     \nonumber \end{eqnarray}
for the space $\varepsilon(1)$. Note that, for fixed $a$ 
(with $a = 0$, 1 or 2), the basis 
$\{ \Psi_a(n_{\alpha}) : n_{\alpha} = 0, 1 \ {\rm and} \ 2 \}$ 
spans the regular representation of the cyclic group $Z_3$. Note 
also that, the basis $b_0(1)$ corresponds to the three irreducible 
{\it vector} representations of $Z_3$ while the bases for $a = 1$ 
and $a=2$ correspond to irreducible {\it projective} representations 
of $Z_3$.

\section{Concluding Remarks}

The derivation of the usual (i.e., non-deformed) Lie algebra 
su$_2$ was achieved in Sec. 2 by adapting the 
Schwinger trick$^{45, 47}$ to the case of two deformed 
oscillator algebras corresponding to a coupled pair of 
truncated harmonic oscillators. This constitutes an unsual result 
for Lie algebras. In the context of deformations, we generally 
start from a Lie algebra, then deform it and finally find a 
realization in terms of deformed oscillator algebras. Here 
we started from two $q$-deformed oscillator algebras from which 
we derived the non-deformed Lie algebra su$_2$. 

The polar decomposition of the ladder operators of su$_2$ inherent 
to our derivation of su$_2$ led to the scheme $\{ J^2 , U_r \}$, 
an alternative to the standard scheme $\{ J^2 , J_z \}$ of angular 
momentum theory, a theory familiar to the physicist.  

Some of the known results about MUB's were explored in Sec. 3 
in the framework of angular momentum theory with a special emphasis 
on the unitay operator $U_r$. This shows that the idea of deformations 
(and possibly Hopf algebras), especially for a deformation parameter 
taken as a root of unity, could be useful for investigating MUBs. 
It is also hoped that the so-called Wigner--Racah unit tensors$^{45}$ 
acting on a subspace of constant angular momentum $\varepsilon(j)$ 
and spanning the Lie algebra of the unitary group U$_{2j+1}$ might be 
useful for characterizing the operators $V_{ra}$ of Sec. 3. Furthermore,
it is worth noting that the parameter $r$ in $V_{ra}$ introduces 
a further degree of freedom. In this respect, let us mention that, 
when $2j+1$ is an odd prime number, 
by replacing $r$ in Eq.~(2) by the $m$-dependent 
parameter 
     \begin{eqnarray}
     r(m) = - a \frac{(j+m)^2}{jm}, 
     \quad m \not= 0, 
     \quad a = 0, 1, ..., 2j
     \nonumber \end{eqnarray}
we generate, together with the spherical basis $s(j)$, 
$2j+2$ MUBs for the space $\varepsilon(j)$. This amounts 
in last analysis to redefining the operator $U_r$.

These matters deserve to be further worked out and should be 
the object of a future work.

\section*{Acknowledgements}

The author is grateful to Michel Planat for very interesting 
correspondence (and the suggestion of using Galois fields and 
Galois rings for classifiyng particles and chemical elements).

\end{document}